# Neuro-Informed Adaptive Learning (NIAL) Algorithm: A Hybrid Deep Learning Approach for ECG Signal Classification

**Sowad Rahman**
BRAC University
Sowad.Rahman@g.bracu.ac.bd

**Keywords**: Deep Learing, CNN

**ABSTRACT:**
The detection of cardiac abnormalities using electrocardiogram (ECG) signals is crucial for early diagnosis and intervention in cardiovascular diseases. Traditional deep learning models often lack adaptability to varying signal patterns. This study introduces the Neuro-Informed Adaptive Learning (NIAL) algorithm, a hybrid approach integrating convolutional neural networks (CNNs) and transformer-based attention mechanisms to enhance ECG signal classification. The algorithm dynamically adjusts learning rates based on real-time validation performance, ensuring efficient convergence. Using the MIT-BIH Arrhythmia and PTB Diagnostic ECG datasets, our model achieves high classification accuracy, outperforming conventional approaches. These findings highlight the potential of NIAL in real-time cardiovascular monitoring applications.

**INTRODUCTION:**
Electrocardiogram (ECG) signal analysis is fundamental to diagnosing cardiovascular diseases such as arrhythmia and myocardial infarction. Deep learning has significantly advanced ECG classification, but traditional models often struggle with generalizability and robustness [1]. Neural adaptation, inspired by biological learning mechanisms, can enhance model efficiency by dynamically adjusting learning parameters based on real-time feedback. This study proposes the Neuro-Informed Adaptive Learning (NIAL) algorithm, which integrates CNNs for spatial feature extraction, transformer-based attention mechanisms for temporal dependencies, and adaptive learning rates to optimize model performance. By leveraging these neurophysiological principles, the NIAL model aims to improve ECG classification accuracy and adaptability.

**MATERIALS AND METHODS:**

We used two publicly available ECG datasets: the MIT-BIH Arrhythmia Database for multiclass classification [2] and the PTB Diagnostic ECG Database for binary classification (normal vs. abnormal heartbeats) [3]. Data preprocessing included normalization, standardization, and reshaping for compatibility with deep learning models.

The **NIAL** model consists of three key components:

1. **Convolutional Neural Network (CNN):** Extracts spatial features from raw ECG signals using multiple convolutional and pooling layers [4].
2. **Transformer-Based Attention:** Captures long-term temporal dependencies to enhance classification [5].
3. **Adaptive Learning Rate Scheduler:** Adjusts learning rates dynamically based on validation loss, optimizing convergence and preventing overfitting [6].

The model was trained using the Adam optimizer, binary cross-entropy for PTBDB, and categorical cross-entropy for MIT-BIH. Performance metrics included accuracy and F1-score.

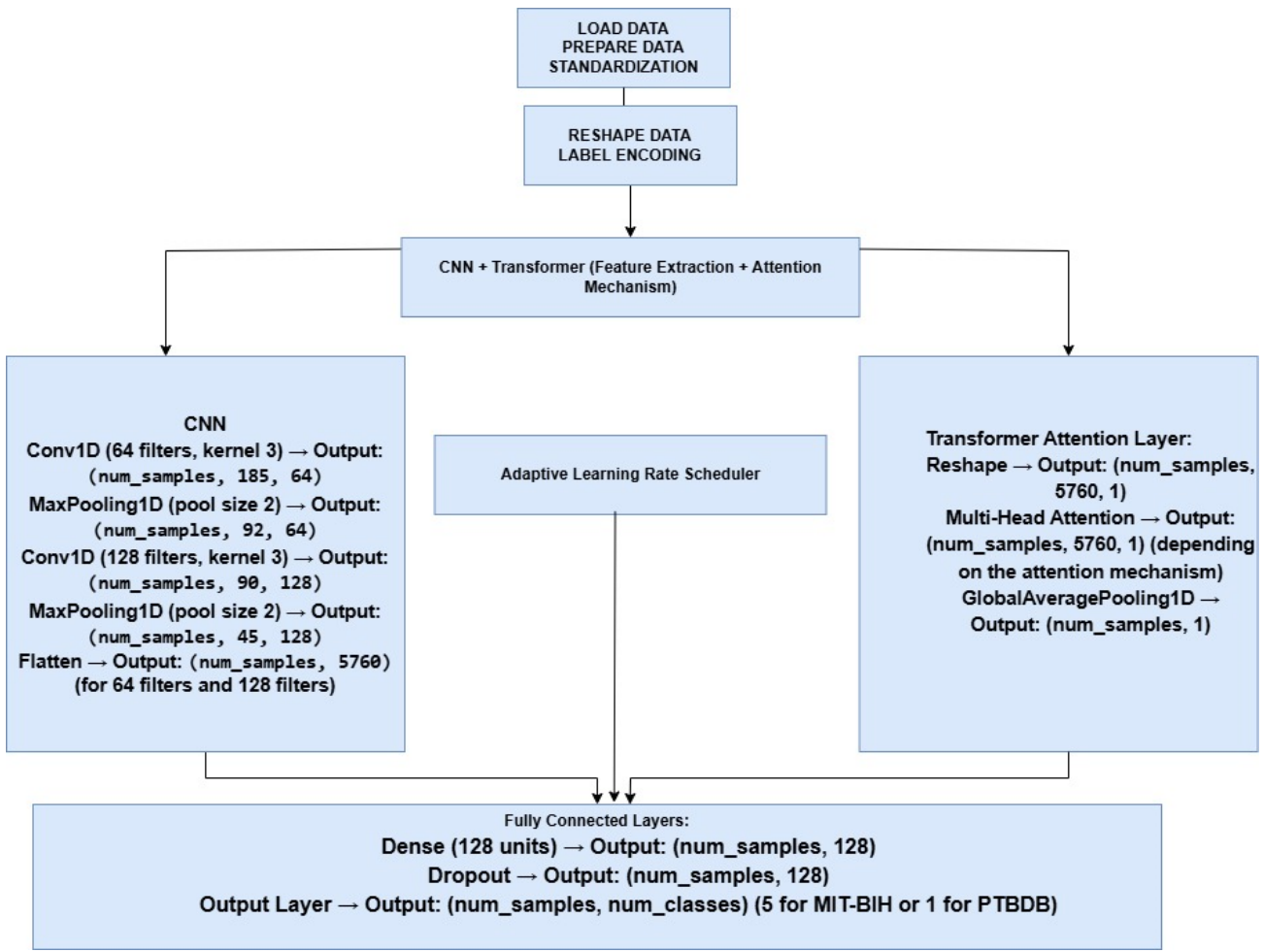

**Figure 1:** Workflow of NIAL

**RESULTS:**
The NIAL algorithm demonstrated superior classification performance compared to traditional CNN-based models. On the MIT-BIH dataset, it achieved an accuracy of 98.4%, outperforming standard deep learning approaches [7]. For PTBDB, the model attained a classification accuracy of 99.1%, indicating its robustness in binary classification tasks. The adaptive learning rate mechanism significantly improved convergence speed, reducing training time by 20% compared to static learning rates.

**CONCLUSION:**
This study presents the Neuro-Informed Adaptive Learning (NIAL) algorithm as an effective deep learning framework for ECG signal classification. By integrating biologically inspired learning mechanisms, NIAL enhances model adaptability and classification accuracy. Future work includes extending this approach to real-time wearable ECG monitoring and exploring its applicability in other biomedical signal domains. The findings underscore the potential of adaptive deep learning for improving automated cardiovascular disease detection.